\documentclass[12pt,onecolumn]{article}

\usepackage[top=1in, bottom=1in, left=1in, right=1in]{geometry}\usepackage[utf8]{inputenc}

\usepackage{lineno}
\usepackage{gensymb}
\usepackage{amssymb}
\usepackage{amsmath}
\usepackage{setspace}
\usepackage{graphicx,float,wrapfig, caption, nicefrac}
\usepackage{placeins}
\usepackage{soul}
\usepackage[dvipsnames]{xcolor}

\title{Droplet tilings in precessive fields: hysteresis, elastic defects, and annealing}

\begin{document}

\maketitle

\noindent
\author{Anton Molina$^{1,2}$, Manu Prakash$^{2 \ast}$

\noindent
\normalsize{$^{1}$Department of Materials Science and Engineering, $^{2}$Department of Bioengineering}

\noindent
\normalsize{Stanford University, 450 Serra Mall, Stanford, California 94305}

\noindent
\normalsize{$^\ast$To whom correspondence should be addressed; E-mail:  manup@stanford.edu}\\
}

\newpage
\vspace{-7pt}
\footnotesize
\begin{abstract}
Two-component Marangoni contracted droplets can be arranged into arbitrary two-dimensional tiling patterns where they display rich dynamics due to vapor mediated long-range interactions. Recent work has characterized the centered hexagonal honeycomb lattice, showing it to be a highly frustrated system with many metastable states and relaxation occurring over multiple timescales [Molina \textit{et al., PNAS}, 2021, \textbf{18}, 34]. Here, we study this system under the influence of a rotating gravitational field. High amplitudes are able to completely disrupt droplet-droplet interactions, making it possible to identify a transition between field-dominated to interaction-dominated regimes. The system displays complex hysteresis behavior, the details of which are connected to the emergence of linear mesoscale structures. These mesoscale features display an elasticity that is governed by the balance between gravity and long-ranged vapor-mediated attractions. We find that disorder plays an important role in determining the dynamics of these features. Finally, we demonstrate the ability to anneal the system by progressively reducing the field amplitude, a process that reduces configurational energy compared to a rapid quench. The ability to manipulate vapor-mediated interactions in deliberately designed droplet tilings provides a novel platform for table-top explorations of multi-body interactions.
\end{abstract}

\newpage
\section{Introduction}

Geometry is fundamental in determining the properties of materials. Until quite recently, materials science has been constrained to understanding the behavior of materials whose geometry is provided by nature. Deliberately designed lattices have emerged to provide engineered alternatives\cite{nisoli2013}. Most generally, lattice systems are a set of discrete degrees of freedom confined to interact according to some spatial pattern. The details of interaction will vary significantly based on a given physical realization whereas the geometric contribution is universal, for example in topological and geometrically frustrated materials \cite{coulais2023}. A number of distinct physical realizations of deliberately designed lattice systems have been developed such as magnetic nanoscale islands \cite{cowburn2000, wang2006, wang2007, mengotti2010, kohli2011, budrikis2011, heyderman2022, zhang2023}, trapped colloidal particles \cite{brunner2002, libal2006, libal2012, ortiz2016, tierno2023}, engineered proteins \cite{baker2015}, and, more recently, evaporating two-component droplets \cite{molina2021}. 

External fields are a common tool for manipulating the behavior of interacting many-body systems. This is especially true for athermal systems where dynamics are otherwise arrested \cite{wang2006, wang2007}. The coupling of external fields to disordered systems can give rise to hysteresis \cite{sethna1993, kohli2011}. While hysteresis has been proposed as means for reducing configuration energy in disordered systems \cite{zarand2002}, its application has not been always successful \cite{zapperi2004}. Hysteresis itself can encompass a variety of phenomena ranging from return point \cite{sethna1993}, multiperiodic \cite{deutsch2003}, and non-convergent memory \cite{hovorka2008}. The qualitative features of hysteresis depend on lattice geometry \cite{sabhapandit2002, deutsch2005, libal2012} where they might correspond to mesoscale phenomena such as domain wall propagation \cite{lyuksyutov1999, domenichini2019} or magnetic monopoles \cite{mengotti2010}. Recent work suggests that such structures can be topologically constrained if the underlying lattice is so-designed \cite{zhang2023}. System which enable the study of hysteresis subject to engineered geometric constraints will provide inspiration for the design of new computational substrates \cite{cowburn2000, loehr2016, arva2019, jensen2020, penty2021, mohseni2022, heyderman2022}.

Liquid droplets are a discrete form of macroscopic matter whose composition can impart an incredible diversity of properties \cite{zarzar2022}, making them ideal candidates to enable table top-top \cite{blonder1984} explorations of many-body phenomena \cite{zarzar2020, molina2021, saenz2021}. More specifically, evaporating droplets are a class of droplet that can interact with each other over long distances. This long range interaction is mediated through the vapor phase where molecular signals propagate through diffusion \cite{houghton1933, pieroban2010}. Formally, systems with an interparticle potential decaying as $1/r^{\beta}$ with $\beta < D$ in dimension $D$ are regarded as long-ranged \cite{LRIBook}. The vapor profiles describing the concentration of the volatile component $\phi$  established by evaporating spherical droplets follow a $\phi \sim 1 / r$ scaling \cite{eggers2010}, satisfying this criteria in $2D$. Previous work on cooperative evaporation of sessile droplets has shown that shielding effects lead to non-additive evaporation \cite{schafle1999, lacasta1998}, extended droplet lifetimes \cite{carrier2016, pandey2020, edwards2021}, and evaporation induced flows (i.e. the coffee ring effect). Recent work has proven successful in understanding several of these aspects, notably evaporation induced flows \cite{wray2019, wray2020, wray2021, stone2021}. 

In contrast to sessile droplets, two-component droplets composed of the right miscible liquids will feature the absence of an additive pinning force \cite{benusiglio2018} and a capacity for motion in response to external sources of vapor \cite{cira2015, benusiglio2017}. This capacity for motion arises from evaporation-induced concentration gradients within the droplet itself which creates a radial surface tension gradient that stabilizes the droplet against spreading. Introducing an external vapor source will break symmetry along the droplet's circumference and produce a net force in the direction of the external vapor source. Thus, the vapor field can be identified as the potential governing droplet dynamics such that $F_{net} \sim \nabla \phi$.

This result has recently been shown to extend to large arrays of evaporating two-component droplets \cite{molina2021}. In this many-droplet regime, phenomena associated with cooperative evaporation such as shielding effects become significant. Due to the motility of two-component droplets, studies on many-droplet evaporation require the presence of a potential well to prevent coalescence. In our past work, physical realizations of a potential well consist of a thin layer of hydrophobic material \cite{molina2021}. Nearly any two-dimensional shape and tessellations thereof can be fabricated to function as a confining potential-well. In this light, designed droplet lattices represent a complement to other deliberately designed lattice systems that is instead based on the physics of evaporation and diffusion. To date, two-component many-droplet evaporation has only been studied in detail for the centered hexagonal honeycomb lattice where it was shown that the vapor-mediated many-body potential gives rise to droplet dynamics that share features with other complex systems, such as frustration, the presence of many metastable states, and relaxation over multiple timescales. However, the use of an external gravitational field to manipulate many-droplet droplet dynamics has not been demonstrated.

Here, we study the dynamics of two-component droplets interacting via cooperative evaporation subject to a time-varying gravitational field. The droplets are confined to a centered hexagonal honeycomb lattice. In the absence of an external field, this system is frustrated and exhibits multi-timescale relaxation. We first build on previous work and provide a theoretical description of this system and identify that droplet size serves as a field-coupling parameter. We characterize the phase behavior of the system experimentally and complement our observations with numerical simulations. We define an order parameter analogous to magnetization and identify a transition between field-dominated and interaction-dominated regimes. We observe hysteresis in this system characterized by limited return point memory. In the transition region, disorder disrupts this type of memory. Additionally, disorder plays an important role in determining the formation of linear elastic defects as the field strength is reduced.  Finally, we show experimentally that field-driven annealing can be used to reduce the frustration of this system. This is the first study concerning the behavior of cooperative evaporation subject to an external field. 

\section{Theory and numerical framework}

Here, we build on previous theoretical and experimental results to construct a numerical simulation framework for our system. The simulation will allow us to provide context to our experimental results. In specific, we use this simulation to study the effects of noise and length scale on non-additive vapor mediated interactions. 

We begin by describing the basic features associated with the evaporation of two-component Marangoni-contracted droplets. It is well known that evaporation at the edge of a droplet is greater than at its center. Thus, when a droplet is composed of two miscible fluids, where one fluid has a higher surface tension and lower vapor pressure (e.g. water and propylene glycol), an effective contact angle $\theta_{eff}$ is observed where otherwise complete wetting is expected. This effective contact angle is due to a Marangoni flow from the droplet apex to its boarder that stops the otherwise individual components of the droplet from spreading \cite{cira2015}. Across a wide range of concentrations, $\cos(\theta_{eff})$ has been shown to be a linear function of humidity: $cos(\theta_{eff}) = mRH + b$ \cite{cira2015, benusiglio2018}. This humidity dependent contact angle gives rise to an absence of pinning and the ability to move under external humidity gradients. 

The behavior of individual two-component droplets subject to gravitational force has been studied previously in a series of inclined plane experiments \cite{benusiglio2018}. Whereas typical sessile droplets experience an additive pinning force which restricts their motion below a certain critical angle, two-component droplets experience a negligible pinning force such that the capillary number $Ca$ is proportional to the Bond number $Bo$ according to $Ca = Bo sin(\alpha)$ (Fig. \ref{fig:figure1}B). This relationship can be derived by equating drag with an external force such as gravity. A crude approximation of the drag force acting on the contact line in the direction of the droplets motion is given by $ F_{drag} = \frac{\pi R \eta ln(b/a) }{\theta}\dot{\vec{r}}$. Where $\dot{\vec{r}}$ describes the velocity of the contact line normal to the plane of motion and $A = \theta / ln(b/a)$. This term is a numerical factor describing the singular dissipation in the moving droplet \cite{deGennes1985}. The terms $b$ and $a$ describe the length scales associated with the details of the physical process of viscous dissipation as the droplet moves. Typically, $b$ represents the macroscopic length scale on the order of droplet radius while $a$ is more difficult to specify. For sessile droplets $a$ might be on the order of the size of a molecule of water. However, droplet motion on a wet or lubricated surface gives a significantly longer cut-off length \cite{keiser2017, benusiglio2018}. A force balance between a gravitational force $F_{g} = 4 \pi R^{3} \rho g sin(\alpha) u(t) / 3$, gives an expression for $\dot{\vec{r}}$:

\begin{equation}
\label{eqn:time_vary_gravity}
    \dot{\vec{r_{i}}} = \frac{a^2 \rho g A}{3 \eta} R^{2} sin(\alpha) u(t)
\end{equation}

\noindent
The addition of $u(t)$ accounts for a time varying applied force. Inspection of eqn. \ref{eqn:time_vary_gravity} allows for identification of the quantity $R \eta$ as the viscous dissipation. Additionally, this expression has been adapted to the multi-body context by including subscripts representing the $i^{th}$ droplet. Since the experiments described in this work concern a rotating gravitational field characterized by a frequency $\omega$, we take  $u(t) = e^{i 2 \pi \omega t}$. Importantly, this term reflects the observation that $Ca = Bo sin(\alpha)$ such that motion proceeds without an additive pinning force. It was noted that above a certain value of $Bo sin(\alpha)$, the receding contact line deforms and eventually enter a regime where they emit smaller droplets. Similar behavior has been reported for single-component droplets on inclined planes \cite{legrand2005}. The effect of gravity on evaporation that remain poorly understood \cite{kim2017, timm2019}. However, these effects are consequential for the internal Marangoni flow \cite{du2015, saenz2017} which underlies motility in two-component droplets. 

The behavior of individual and pairs of two-component Marangoni-contracted droplets subject to external humidity gradients has also been studied previously \cite{cira2015, benusiglio2017}. A two-component droplet $i$ can move in response to a humidity gradient established by a neighboring droplet $j$ separated by a distance $r_{ij}$ (Fig. \ref{fig:figure1}A). The magnitude of this force is given by $F_a = \gamma R_{i} m f(R_{j}, r_{i})$. Here, $m$ is the slope describing the linear relation between $\theta_{eff}$ and $RH$. The function $f(R_{j}, r_{ij})$ has previously been identified as the gradient in the vapor concentration established by the $j^{th}$ droplet. Thus, $f(R_{j}, r_{i}) = \nabla \phi(R_{j}, r_{ij})$ and scales with $1/r_{ij}^{2}$ \cite{molina2021}. The vapor concentration of a lens-shaped droplet with contact angle $\theta \sim 0^{\circ}$ can be described by \cite{eggers2010}: 

\begin{equation}
\label{eqn:vapor_field}
    \phi(R_{j}, r_{ij}) = 
    \begin{cases}
    1 & r_{ij} < R_{j} \\
    \frac{2}{\pi} \arcsin{(\frac{R_{j}}{r_{ij}})} & r_{ij} \geq R_{j}
    \end{cases}
\end{equation}

\noindent
The low-contact angle limit is valid since the high-humidity associated with a dense array of droplets is high. This situation can be conceptualized as individual droplets evaporating into a global "super drop" composed of the system's shared vapor  \cite{carrier2016}. This expression has been used to describe the evolution of droplet lifetimes in the context of sessile multi-droplet evaporation  and more recently in vapor-gradient driven dynamics of two-component Marangoni-contracted multi-droplet systems \cite{molina2021}. In the later case, it was necessary to introduce a fitting parameter $\xi$ representing number of $N^{th}$-nearest neighbor interactions to describe non-additive shielding effects associated with cooperative evaporation. 

We can combine eqn. \ref{eqn:time_vary_gravity} and \ref{eqn:vapor_field} into a single expression to describe the dynamics of a many-droplet system subject to external driving. The final, non-dimensional expression is obtained after dividing by the characteristic velocity $\gamma / \eta$: 

\begin{equation}
\label{eqn:eqn_combined}
    \dot{\vec{r}} = 
    \frac{2 mRH}{\pi} \nabla \sum^{N}_{i} \sum_{j}^{\mathcal{N}(\xi)} \arcsin{(\frac{R_{j}}{r_{ij}})} + 
    \frac{a^2 \rho g A}{3 \gamma} \sum^{N}_{i}  R_{i}^{2} sin(\alpha) e^{i 2 \pi \omega t}
\end{equation}

\noindent
This expression can be used to construct a numerical simulation. In addition to the basic empirical observations described above, we make several further approximations. First,  individual droplets with radius $R_{i}$ are modeled as hard disks. That is the droplets are never non-spherical, which might have consequences for evaporation dynamics \cite{kim2017, timm2019, du2015, saenz2017}. Further, and more significantly, this model ignores the diffusion timescale in the system: vapor profile are established instantaneously to a spatial extent determined by $\xi$. Finally, a potential well is defined to enforce the constraint of a hydrophobic boundary (see Materials and Methods).  This treatment ignores the possibility of contact line pinning or any friction associated with the hydrophobic boundary. 

\section{Results}

\begin{figure}[h]
\centering
  \includegraphics[height=11cm]{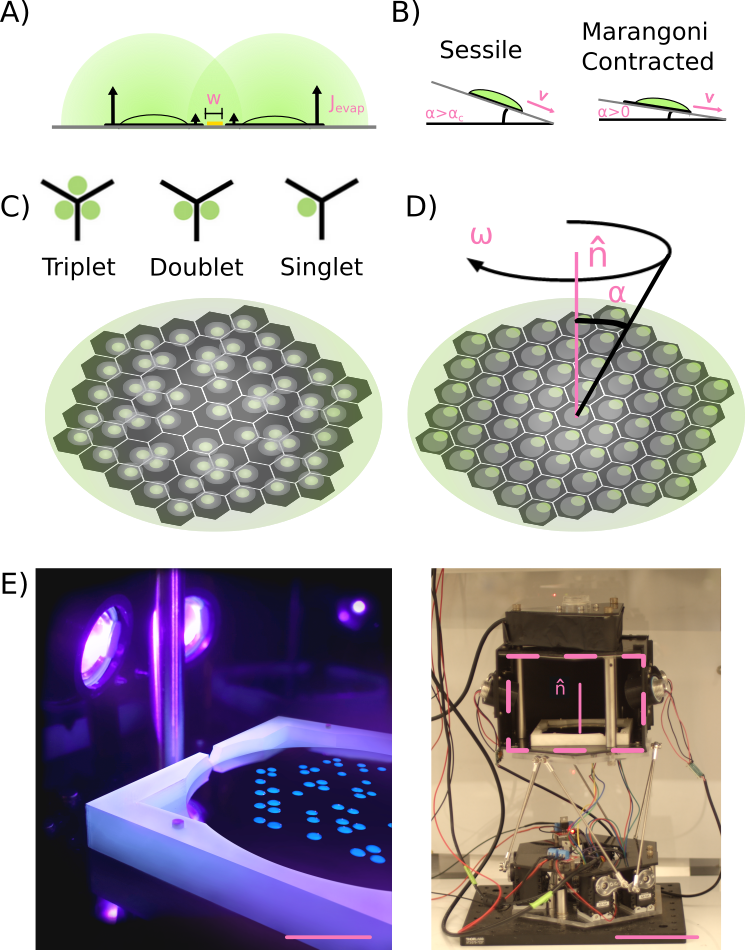}
  \caption{Experimental setup.
  (A) schematic showing a pair of two-component Marangoni-contracted droplets separated by a hydrophobic barrier of width $w$ interacting through evaporation.
  (B) schematic showing that droplet motion occurs without an additive pinning force at critical angle $\alpha_{c}$ for Marangoni-contracted droplets. 
  (C) In the absence of an external gravitational force, a many droplet system will form fixed structures at the vertex of the lattice. Cutaway shows the types of vertex structures that can form on the hexagonal honeycomb lattice.
  (D) A rotating gravitational force of sufficient strength can break these vertex structures and drive droplet motion
  (E) Experimental realization of a frustrated hexagonal honeycomb lattice where droplets are illuminated under UV illumination. Scale bar corresponds to 23 $mm.$ The cutaway shows the system placed inside of an enclosed chamber mounted atop of a Stewart platform. Scale bar corresponds to 100 $mm$.}
  \label{fig:figure1}
\end{figure}

We study the centered hexagonal honeycomb lattice with $N=61$ two-component droplets with composition 70\% water and 30\% propylene glycol. As noted, the relaxation behavior of this system in the absence of an external field was studied previously \cite{molina2021}. The system organizes itself to form vertex structures, where each vertex can be categorized according to the number of occupying droplets (Fig. \ref{fig:figure1}C). The locally preferred structure corresponds to maximum droplet occupancy. We subsequently refer to this structure as a "triplet", due to the 3-droplet maximum occupancy permitted by the geometry of the hexagonal honeycomb lattice. This structure minimizes evaporation and extends droplet lifetimes \cite{wray2019, wray2021}. The arrangement of these structures will break symmetry in the global super drop, giving rise to complex dynamics. The ground state of this system can be defined as the state in which the number of locally preferred triplet structures is maximized. It was found that long-range interactions and boundary effects lead to a highly frustrated system. We expect that a rotating, external field can disrupt these interactions and effectively melt the system (Fig. \ref{fig:figure1}D). In this work, the field is imposed by mounting an enclosed chamber containing the droplet array onto a Stewart platform (Fig. \ref{fig:figure1}E, SI Video 1) \cite{Stewart1965, patel2018}. The aim of the present study is characterize the behavior of this system subject to an external field and to understand how an external field can disrupt a disordered, elastic system governed by vapor-mediated interactions.

\begin{table}[h]
\centering
\small
  \caption{\ Physical properties of two-component droplets}
  \label{tbl:example1}
  \begin{tabular*}{0.48\textwidth}{@{\extracolsep{\fill}}lll}
    \hline
    Quantity (units) & Value & Source\\
    \hline
    $\rho$ ($\frac{kg}{m^3}$) & 1030 & Ref. \cite{khattab2017} \\
    $\eta$ ($\frac{kg}{m s}$) & 12.0 & Ref. \normalsize{\cite{george2003}} \\
    $\gamma$ ($\frac{N}{m}$) & 41.4 & Ref. \cite{khattab2017} \\
    $\theta$ ($^{\circ}$) & 10 & Ref. \cite{benusiglio2018}\\
    $m$ & $3$ $10^{-4}$ & Ref. \cite{benusiglio2017}
  \end{tabular*}
\label{table:physParam}
\end{table}

\subsection{Transition from interaction- to field-dominated behavior}

\begin{figure*}
 \centering
 \includegraphics[height=9.0cm]{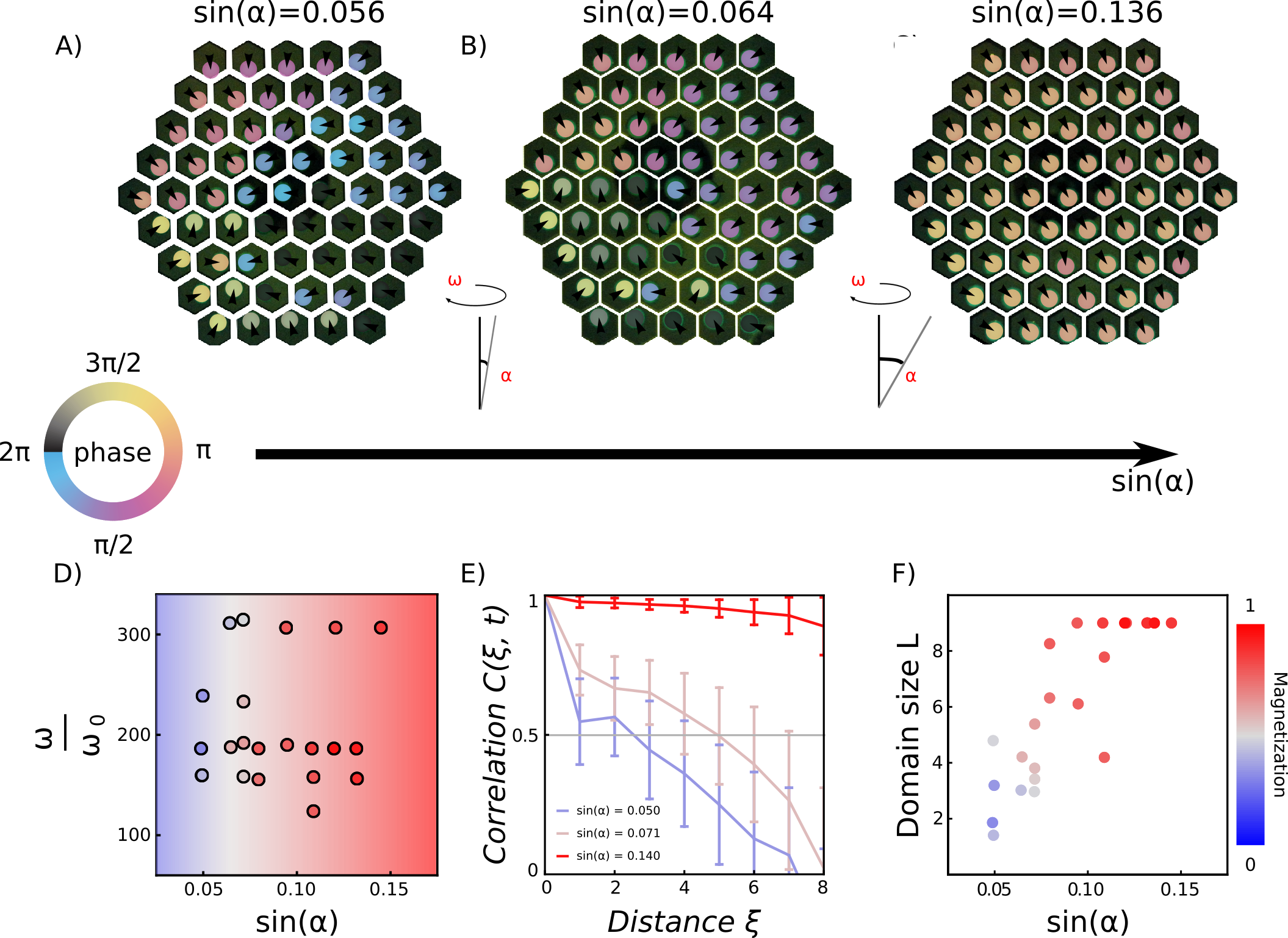}
 \caption{Phase behavior of frustrated droplet system. 
 (A) interaction-dominated phase at $sin(\alpha) = 0.049$ and $\omega / \omega_{0}$ = 186, all droplets are stationary with a mix of phase. 
 (B) transition phase at $sin(\alpha) = 0.064$ and $\omega / \omega_{0}$ = 311, there is a mix of stationary and mobile droplets and a mix of phase.
 (C) field-saturated phase at $sin(\alpha) = 0.136$ and $\omega / \omega_{0}$ = 229, all droplets are mobile and in phase.
 (D) phase diagram in the $\omega / \omega_{0}$ - $sin(\alpha)$ plane obtained from experimental data. Here, we identify three regions on the basis of their magnetization: field-saturated, transition, and interaction-dominated.
 (E) equal-time correlation function for representative experiments from each region. Error bars show the standard deviation in the correlation function at 10 randomly sampled times. Gray line at 1/2 shows correlation cut-off used to define domain size.
 (F) domain length scale as determined from equal-time correlation function as a function of $sin(\alpha)$.
 }
 \label{fig:figure2}
\end{figure*}

We begin by exploring the phase diagram defined by the field amplitude $sin(\alpha)$ and frequency $\omega / \omega_{0}$. Data was collected within a region of this plane defined by the range $sin(\alpha) = 0.048-0.145$ and $\omega / \omega_{0} = 157-314$. These values represent the amplitude and frequency limits of our homemade Stewart platform. Our first set of experiments consist of droplets with average radius $<R>=0.21a$. On average, droplet diameters are reduced by only 1.9\% at the end of an experiment (SI Fig. 1). Inspection of the data allows us to identify three distinct states: interaction-dominated, transition, and field-dominated (Figure \ref{fig:figure2}A-C, SI Video 2). In the interaction dominated state, droplet motion is restricted but non-zero. Due to the absence of an additive pinning force, the external field is still able to affect dynamics even at extremely shallow angles. Interestingly, this type of dynamics appears to suppress the long-time scale rearrangements towards the center of the global vapor cloud that occur when $sin(\alpha)=0$ \cite{molina2021}. Coloring each droplet according to its phase, shows a partial cancellation of phase as droplets form local vertex structures (Fig. \ref{fig:figure2}A). As the field strength increases, so too does droplet mobility. Droplets in the interior of the system display a high degree of mobility allowing them to explore multiple configurations. While local vertex structures are either suppressed or transient. More specifically, droplets that are two to three lattice sites interior from the boundary show restricted paths of motion oriented towards the center of the system, suggesting that their motion is influenced by the global vapor cloud (Fig. \ref{fig:figure2}B). Eventually, the external field can completely saturate droplet interactions and lead to a state characterized by a high degree of phase coherence (Fig. \ref{fig:figure2}C).

To better characterize the phase behavior of this system we define an order parameter. Observation of phase delay and synchronization behavior suggests an approach grounded in the Kuramoto model of coupled oscillators. Alternatively, observation of field saturation and domains of aligned droplets suggests the perspective of magnetism. Since the droplets lack an intrinsic frequency, we pursued the latter approach. We define an order parameter $M = \frac{1}{N} \sum_{i}^{N} s_{i}$, where $s_{i}$ is the unit vector with the orientation $\psi_{i}$ of the $i^{th}$ droplet. Analogously with magnetism, this quantity  characterizes the ability of the external field to saturate the "spin" orientation: $M$ can range from 0 to 1; when $M=1$, all droplets are oriented in the same direction. Due to the finite size of the system we expect $M>0$ since complete cancellation of spins will be unlikely. It is worth noting, that this order parameter is insensitive to deformations in droplet shape.

We now compute the time-averaged magnetization $<M>_{t}$ for 21 different experimental realization in the region of the $sin(\alpha)$ and $\omega / \omega_{0}$ plane described above (Fig. \ref{fig:figure2}D). Within this region, we find that the transition from low to high values of $M$ that occurs with increasing $sin(\alpha)$ and is comparatively insensitive to $\omega / \omega_{0}$. Further, we characterize the spatial cohesion of the system through this transition by calculating the equal-time correlation function with respect to droplet orientation (see Materials and Methods). This quantity contains information about how much neighboring droplets influence distant neighbors subject to increasing values of $sin(\alpha)$ (Fig. \ref{fig:figure2}E). The slow decay in the correlation function can be attributed to the long-range interactions. Weakly interacting systems typically have exponentially decaying correlations. Cohesion means that below a certain length $L$, droplets will share the same orientation. The equal-time correlation function will vary from one at $\xi=0$ to zero at $\xi >> L$, and will cross $1/2$ at a characteristic domain size $L$ \cite{sethna2006}. We see a step, continuous increase in $L$ as the field strength is increased, with droplet orientation saturating at field strengths $sin(\alpha)>0.1$ (Fig. \ref{fig:figure2}F).

To contextualize these experimental observations, we perform a series of numerical simulations. The physical parameters defining eqn. (\ref{eqn:eqn_combined}) are estimated from values previously reported in literature and are collected in Table \ref{table:physParam}. We use the lattice dimension $a=8$ $mm$ as a characteristic length scale from which we can calculate the characteristic timescale $t_{0} = a \eta / \gamma$ (2.3 $ms$). We first explore how the phase diagram depends on droplet size assuming uniform droplet size of $R = 0.2a$ and $R = 0.25a$ for varying number of neighbor interactions $\xi=1-4$ (SI Fig.  2). We can identify the same three phases observed in experiment. Additionally, we identify a fourth state where the driving period is too fast for $M$ to saturate, giving the transition region a dependence on $\omega$. However, these frequencies are too fast for us to access experimentally in our current platform. We can see that increasing $\xi$ has the effect of broadening the transition region characterized by intermediate values of $M$. Further, the fully magnetized state ($M=1$) can only be realized at very large values of $sin(\alpha)$ or for low values of $\xi$. This can be attributed to strong finite-size effects on the $N=61$ lattice: droplets at the edge experience an attraction to the concentrated vapor at the center, requiring a relatively high value of $sin(\alpha)$ to disrupt this interaction. To understand the role of droplet size, we consider the series of phase diagrams obtained for the system composed of smaller droplets ($R=0.2a$). The qualitative features of the phase diagram are similar; however, the magnetized phase is shifted towards higher values of $sin{(\alpha)}$ and is not observed for high frequency driving. This can be explained by interpreting droplet size as a field coupling term where the influence of gravity scales with $R^{2}$. So, larger fields are required to drive smaller droplets. Finally, we see how the phase space depends on noise (SI Fig. 3, SI Video 3). We consider noise in initial conditions and in droplet size. Introducing noise (5\%) in the initial positions has little effect on the phase diagram. However, considering the variance of $M$ in the $sin(\alpha)$-$\omega / \omega_{0}$ plane, we see a high variance appear in the region where the external field is unable to exert an influence over dynamics. Introducing noise (10\%) in droplet radius ($<R>=0.225a$), has the effect of broadening the transition region. Unlike, noise in initial conditions, noise in droplet radius is nonzero across the entire $sin(\alpha)$-$\omega / \omega_{0}$ plane. Interestingly, the region of greatest variance coincides with the transition region. 

Within the region of the $sin(\alpha)$-$\omega / \omega_{0}$ plane studied, we observe the same qualitative states both in experiment and and numerical simulation. In particular, field-saturation of the droplets occurs at much smaller values of $\sin(\alpha)$ than expected. We have found that interaction length scale, noise in droplet size or initial position are insufficient for explaining this discrepancy. We suspect that cooperative evaporation effects might be responsible for this discrepancy in reducing the saturation field.

\subsection{Hysteresis and mesoscale elastic structures}

We now consider the evolution of the order parameter with time. In experiment, we observe that M is a quasi-periodic function of time characterized by a large amplitude $|M|$, particularly in the transition- and interaction-dominated regimes (SI Fig. 4). The value $|M|$ corresponds to the spread in droplet orientations $\psi_{i}$ within a period. For low driving amplitudes, a spread in $\psi_{i}$ is expected since droplets will arrange themselves into vertex structures. As the driving amplitude is increased, droplets will acquire mobility in an order that depends primarily on their size but also on the details of their neighbors (e.g. their size and shape - insofar as it effects the extent and distribution of vapor). The result is a state characterized by decoherence, phase delay, and domains of spin orientation whose size grows with increasing driving amplitude. At high drive amplitude, where the field can saturate the droplet orientations, the spread in $\psi_{i}$ is minimal and $|M|$ is low. However, we observe that $M$ tends towards lower values over time with larger $|M|$ , a feature particularly pronounced for the highest driving amplitudes. 

\begin{figure*}[h]
\centering
  \includegraphics[height=9cm]{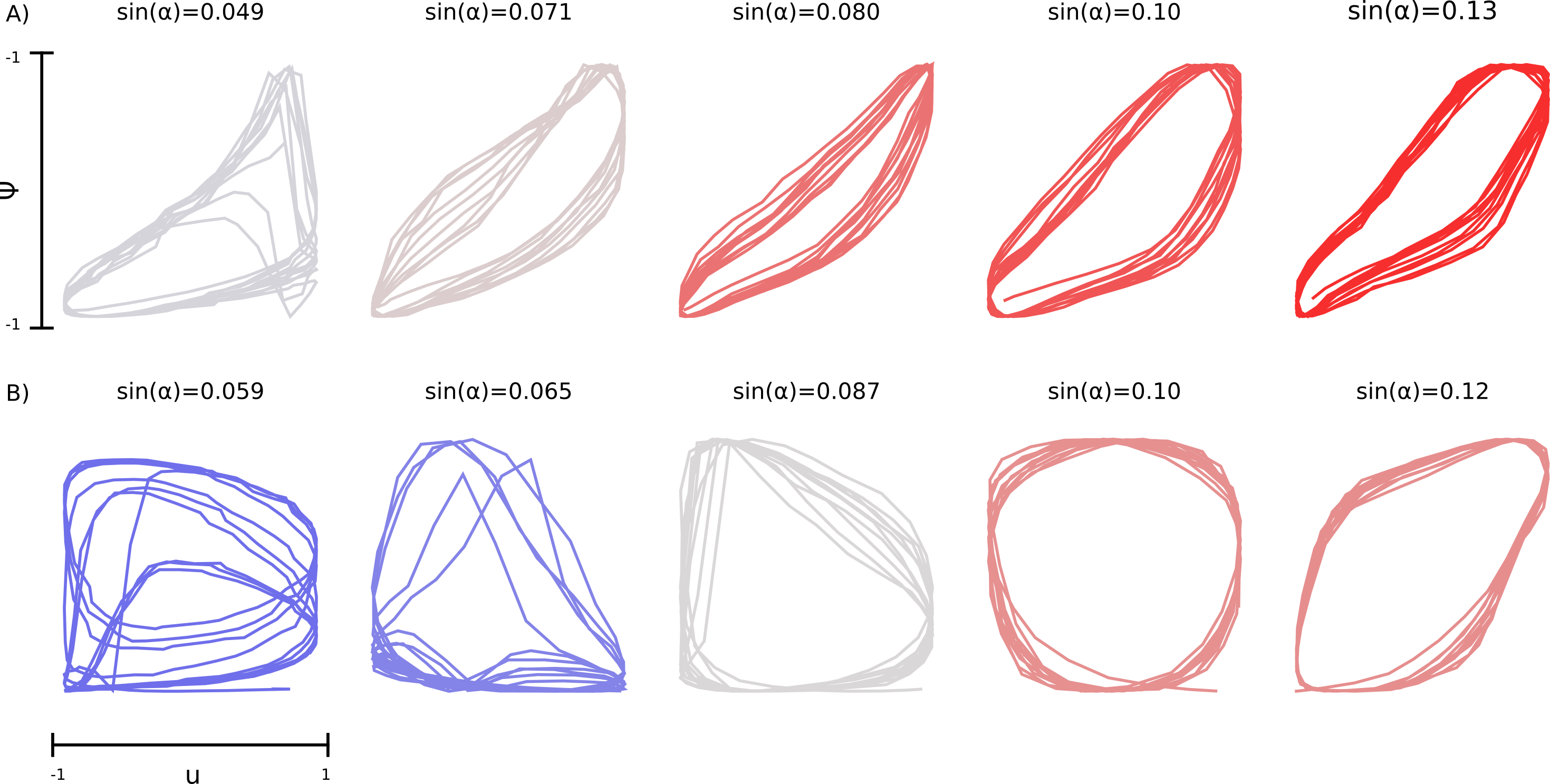}
  \caption{Hysteresis loops measured as a function of $sin(\alpha)$ for systems with average droplet radius (A) $0.2l$ ($\omega / \omega_{0}=155$) and (B) $0.18l$ ($\omega / \omega_{0}=235$). For each loop the horizontal axis is the orientation of the applied field and the vertical axis is magnetization. The loops are colored according to their average magnetization according to the color scheme described in Fig. \ref{fig:figure2}.}
  \label{fig:figure3}
\end{figure*}

The presence of delay and quasi-periodicity suggests that this system might display interesting hysteresis behavior. Most generally, hysteresis describes a particular response of a system to a periodic input \cite{Morris2011}. Here, we define our input as the time-varying orientation $u$ of the rotating gravity vector. We note that the magnitude of the field is given by $sin(\alpha)$. Our output is the the angle $\Psi$ defined as the average of droplet orientations: $\Psi = \frac{1}{N} \sum_{i}^{N} \psi_{i}$. According to these definitions, both the input and output functions can assume values between $-1$ and $1$. Further, a linear relationship between input and output indicates a response with no delay while a circular loop indicates an offset in the collective response of $\pi / 2$. By plotting the input function $u$ against its output $\Psi$, we observe the looping structures characteristic of hysteresis \cite{Morris2011}. Figure \ref{fig:figure3} shows a subset of the measured hysteresis loops at two different rotation frequencies for various values of $sin(\alpha)$. The rotational frequencies vary only slightly ($\omega /\omega_{0}=$ 155 and 235). As noted, this difference in frequency is relatively insignificant for the region of the phase diagram studied experimentally. The more important difference is that these two slices across phase space differ in droplet size ($R=21a$ with 10.1\% noise and $R=18a$ with 8.9\% noise, respectively) (SI Fig. 1). We observe that the systems converge to a well defined limit cycle. However, for the majority of the cycle the loops do not perfectly trace over themselves. Interestingly, there are regions of the limit cycle where the order parameter assumes a type of return memory, in particular when $u = \pm 1$; away from these regions, there is a broader distribution in the value of the order parameter.

\begin{figure*}[h]
\centering
  \includegraphics[height=13cm]{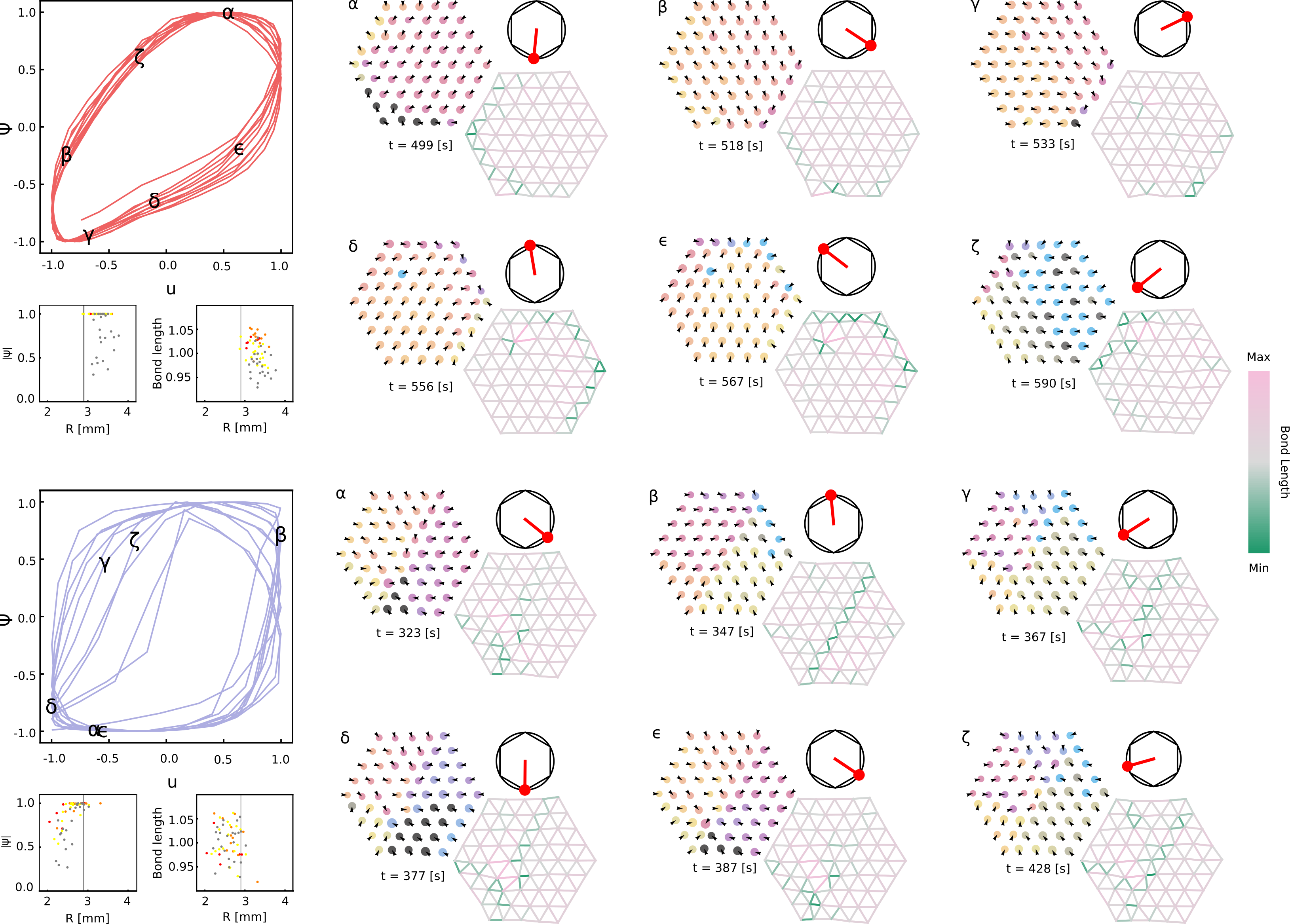}
  \caption{Mesoscale structures emerge during hysteresis depend on disorder. (A) Hysteresis loop obtained for the ordered system $sin(\alpha)=0.095$ and ($\omega / \omega_{0} = 155$). 
  (B) Snapshots of phase and nearest-neighbor bond lengths at intervals (indicated by Greek letters) within a single hysteresis loop. Droplet phase are annotated according to the same scheme as in Figure 2. The orientation of the gravitational field at each snapshot relative to a typical unit cell is indicated by the red arrow.(C) $|psi_{i}|$ and (D) average bond length for individual droplets plotted as a function of droplet size. Data points are colored according to their distance from the center of the system. The gray vertical line indicates mobility cut-off of $0.18a$ at this driving amplitude. (E-H) The same as A-D obtained for the disordered system at $sin(\alpha)=0.095$ and ($\omega / \omega_{0} =235 $)
  }
  \label{fig:figure4}
\end{figure*}

We now connect the microscopic dynamics and the resulting hysteresis loop for two realizations of the system that differ primarily in the size distribution of the droplets ($R=0.21a$; 7.1\% noise and $R=0.16a$;  10.1\%) (Fig. \ref{fig:figure4}, SI Video 4). We call these realizations low and high disorder respectively. It is important to note that there is overlap between the two size distributions (SI fig. 1). This means that the largest droplets in the high disorder realization are nearly the same size as the average droplet size in the low disorder realization. In other words, high disorder is associated with an increased probability of pinning. For clarity, pinning refers not to the contact line pinning often associated with droplets but with elastic pinning associated with elastic, condensed matter systems. Hysteresis loops for these two realizations are obtained at $sin(\alpha)=0.095$ - corresponding to the onset of the transition region. For each realization we look at several snapshots of the system throughout a complete loop. We consider both the phase (introduced previously) and bond length. Here, bond length refers to the distance between nearest neighbor droplets. Finally, we consider the role of droplet size and its position within the lattice on two local dynamic quantities: $|\psi_{i}|$ and standard deviation of average bond length. The former allows us to identify a pinning threshold of $18a$, below which non-boundary droplets show reduced mobility (SI Fig. 5). The latter allows us to identify emergent mesoscale defects.   

The droplet size distribution which arises from experimental noise introduced during the initial droplet placement. results in significant differences in the resulting hysteresis loop (SI Fig. 1). In the low disorder case, we find a loop that converges to a limit cycle characterized by partial return point memory (Figure \ref{fig:figure4}A). In this case, non-boundary droplets respond to the field uniformly and delay is related to the formation of a linear elastic structure at the boundary (Fig.  \ref{fig:figure4}B). The elastic structure is most clearly seen by looking at the evolution of droplet bound lengths. Otherwise, the system presents a coherent domain of spin orientations. Looking at the behavior of individual droplets, we find that only droplets at the boundary show reduced values of $|\psi|$ with no significant relationship on droplet size (Fig. \ref{fig:figure4}C).  Meanwhile, we find that boundary droplets also have the largest variance in bond length (Fig. \ref{fig:figure4}D). A more subtle feature is that the variance in bond length if a function of distance from the center of the system. Again, there is no significant relationship on droplet size. 

When disorder is increased, we obtain a hysteresis loop with reduced overlap (Fig. \ref{fig:figure4}E). This behavior is a result of aperiodicity in the order parameter. In this case, droplets do not respond uniformly to the external field. More, the dynamics of droplets above the cutoff threshold will depend on their proximity to droplets below the threshold. As a result, we observe a linear elastic structure that occurs in the bulk (Fig.  \ref{fig:figure4}F). Again, the elastic structure is most clearly seen by looking at the evolution of bound lengths. Interestingly, the defect acts to divide two regions of otherwise coherent spin orientations and therefore acts as a grain boundary. Examining the behavior of individual droplets shows that reduced droplet mobility, as characterized by $|\psi|$, can occur in the bulk (Fig. \ref{fig:figure4}G) when droplets are below the pinning threshold size. At the same time, the dependence on variance of bond length on distance from system center is removed \ref{fig:figure4}G). In other words, disorder means that structure formation can be initiated away from the boundary. This observation is similar to other disordered systems subject to a rotating field \cite{budrikis2011}.  

Subject to the confines of a potential well, the external field acts to maximize droplet-droplet distances, amounting to an effective repulsive interaction. As a result, we observe a configuration characterized by six-fold coordination, similar to those observed in Coulomb crystals \cite{weitz2015, lippy2021}. To be precise, six-fold coordination in this context means all bond lengths are of nearly equal length. A defect then refers to a bond whose length is significantly different than the average. That defects emerge at the boundary can be explained in two ways: 1) droplets do not feel the pull of their neighbors that would exist beyond the edge; 2) droplets at the edge are subject to a particularly strong attraction to the global vapor cloud whose center of mass lies at the center of the system. In this view, vapor-mediated attractions are a competing force that cause distortions to the ideal lattice structure. The occurrence of defects in the bulk requires is closely associated with disorder. 

Low-field perturbations about stable vertex sites are expected to produce periodic behavior with little dependence on disorder. Quasi-periodic behavior can emerge as a result of three distinct processes. First, phase delay due to droplet size can give rise to dynamical behavior that spans multiple periods. Phase differences can also arise from the mutual influence of small and large droplets on each other. Second, as vapor diffuses across the system, the droplets might move into a different "sphere of influence". Put differently, the details of the droplet's local environment have changed. Additionally, the number of interacting neighbors increases, requiring a strong field to saturate the droplet orientations (SI Fig. 2). These effects will eventually pull and orient droplets towards the system center. This process is shown schematically in SI Fig. 5 and will result in observed hysteresis loops with reduced overlap. Third, since droplets leave behind a trail of their own constituents during motion (SI Video 5), strongly driven  droplets will essentially "age" faster  which might result in a slowing down of their dynamics. Given that this effect is more pronounced in higher amplitude experiments, we suspect this explanation is becomes significant past a certain amplitude threshold.

Comparison with numerical simulation provides some insights into the origin of these dynamical features. If droplet size is uniform (SI Fig. 6), $|M|$ is low. In fact, we observe a sub-harmonic with frequency $6\omega$, corresponding to the six boundary collisions that occur in a centered hexagonal system. $|M|$ broadens only in the interaction-dominated regime where the spread in droplet orientations reflects the formation of vertex structures. Interestingly, we find $M$ varies quasi-periodically with time in the transition regime. When the droplet size distribution is non-uniform (SI Fig. 7), $|M|$ is broad even at high amplitudes. Further, quasi-periodic behavior becomes common, particularly in the transition regime. In both cases, the drift towards lower values of $M$ is not observed, supporting the interpretation that these are ageing affects.

\subsection{Annealing}

\begin{figure}[h]
\centering
  \includegraphics[height=12cm]{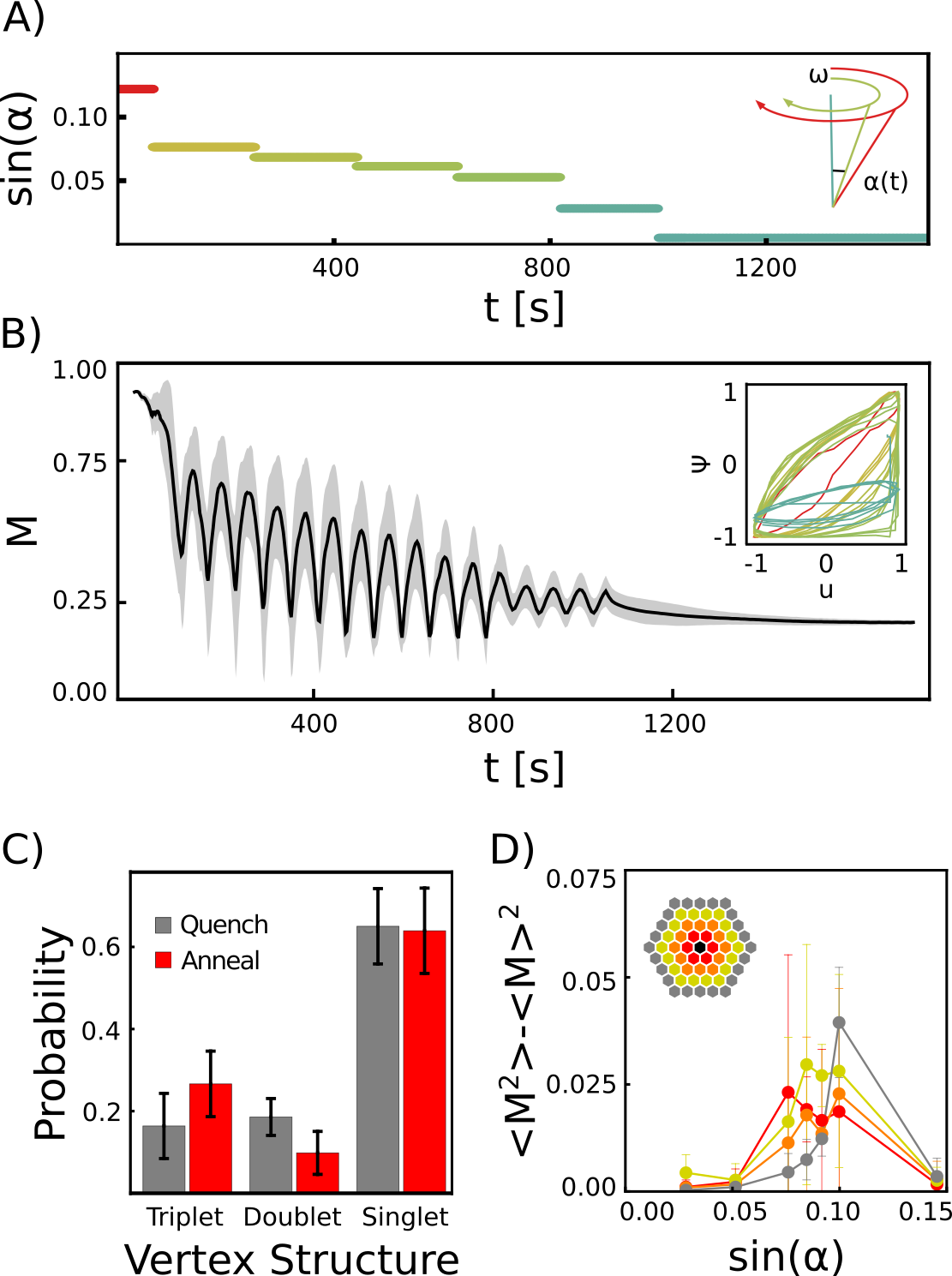}
  \caption{Annealing of a frustrated droplet tiling.
  (A) The annealing schedule used to drive the frustrated system ($\omega / \omega_{0} = 183$).
  (B) Time series showing the evolution of $M$ as the annealing schedule is applied. Variance is for $n=3$ experimental realizations. Inset shows $M$ vs $u(t)$ with hysteresis during annealing.
  (C) Vertex statistics obtained from experimental data comparing a rapid quench ($sin(\alpha)=0$ with a system annealed using the schedule shown in (A). 
  (D) Heat capacity as a function of field strength for increasing distance from the center. Inset shows color corresponding to distance $\xi$ from center.  
  }
  \label{fig:figure5}
\end{figure}

The physical basis of annealing relies on the controlled introduction of randomness during the system's evolution\cite{kirkpatrick1983}. Annealing traditionally relies on randomness from thermal noise to escape local energy traps. However, determining the rate of cooling - the annealing schedule - presents a significant theoretical challenge \cite{hoffmann1990}. A heuristic approach for annealing glassy systems has been suggested \cite{morgenstern1987}, whereby the time spent above a critical temperature is maximized. In this regime, nontrivial spin flips are maximal enabling exploration of the entire energy landscape. It has been shown that a global field can also be used for optimization if coupled to elements with a distribution of responses to the field. A distribution of field couplings can provide the key element of randomness required for optimization \cite{zarand2002}. Here, we apply such an approach to the $N=61$ centered hexagonal honeycomb lattice and see whether we can obtain final configurations that are closer to the idealized ground state described above (Fig. \ref{fig:figure5}, SI Video 6). 

The particular annealing schedule used in this work is shown in Fig. \ref{fig:figure5}A. The majority of time during annealing is spent where $sin(\alpha)$ is in the transition region to maximize non-trivial dynamics. All driving occurs at a frequency $\omega / \omega_{0} = 186$ for three rotations at the specified field strength. The magnetization is shown in Fig. \ref{fig:figure5}B. It can be see that magnetization is steadily reduced with large fluctuations about the mean value until a field strength where interactions dominate and nontrivial spin flips likely can no longer occur. This corresponds to a reduction in the looping behavior identified with hysteresis. 

Finally, we calculate the specific heat defined as fluctuations in the order parameter $<M>^{2}-<M^{2}>$ (Fig. \ref{fig:figure5}C). This quantity describes when individual droplet orientations differ substantially from average behavior. Here, $M$ is calculated for each subset of droplets that are $\xi$ lattice sites from the center. As expected, fluctuations are maximized in the transition region. Interestingly, we find that the occurrence of maximal fluctuations depends also on their spatial distribution within the lattice. Droplets that are closer to the center of the system display a maximum fluctuations at lower values of $sin(\alpha)$. Further, this maximum value obtained increases as distance from center increases. Comparison of statistics describing the distribution of vertex sites obtained after annealing with those obtained after a rapid quench show that the probability a finding a droplet in a triplet configuration increases to 0.27 from 0.16, an increase of 69\% Fig. \ref{fig:figure5}D. Meanwhile, the probability of finding a droplet in a doublet configuration is reduced to 0.1 from 0.19, a reduction of 48\%. We conclude that annealing has the primary effect of converting doublets into triplets.

\section{Conclusions}

We have shown that gravity can be used as external field to perturb the interactions in an array of evaporating two-component droplets. While the resulting dynamics share many similarities with other condensed matter systems, the underlying physics is based on evaporation and remains under explored. Further characterization of the behavior of two-component Marangoni-contracted droplets merits attention. For example, a more detailed understanding of the interplay between motility and internal Marangoni flow in complex vapor profiles established in a multi-droplet array is needed. Even at the scale of isolated droplets, understanding the trailing phenomena observed by the large, strongly-driven droplets observed merits further study. Incorporating the recent theoretical advances related to evaporative flux in a multi-droplet context into a numerical framework is high priority. Incorporation of the diffusion timescale is an important missing ingredient that would likely enable a faithful reproduction of the experimental observations. Recent work emphasizing simplified models amenable to efficient numerical implementation make this a tangible possibility \cite{hartmann2023, stone2021, wray2021}. An accurate and efficient numerical framework would enable the rational development of closed-loop control strategies. Strategies based on deep reinforcement learning have proven fruitful in controlling structure formation in colloidal crystal systems and point the way toward control over more complicated systems\cite{zhang2020}. Stated differently, such an approach could provide new insights to the longstanding question 'what is the optimal annealing schedule?' \cite{hoffmann1990}.

While the presented work has focused on the centered hexagonal honeycomb lattice, we emphasize that a vast design space of tiles and tessellations thereof remain to be explored. Deliberately designed lattices have most thoroughly been explored in realization consisting of nanoscale magnetic islands, so-called artificial ice systems \cite{Sk2020}. Significant progress has been made in mapping exotic geometric arrangements of two-state nanoscale magnets to mesoscale emergent phenomena. A similar effort should be undertaken with systems governed by different physics so as to develop general principles between geometry and dynamics. Such an understanding will be important for the the development of unconventional computing platforms that exploit parallelism and emergent phenomena \cite{arva2019, jensen2020, penty2021, heyderman2022}.

In conclusion, collective evaporation in confined arrays of motile, two-component droplets gives rise to complex emergent dynamics characterized by multiple relaxation timescales. Here, we have shown that gravity can be used as an external field to disrupt these interaction such that a continuous transition between interaction- and field-dominated regimes can be identified. We have shown that hysteresis occurs during driving and specific details are sensitive to disorder in the system. Aperiodic hysteresis is most likely to occur during the transition regime. The micro- and meso-scale dynamics in this region enable the system to access lower energy configurations through an annealing schedule in comparison with a rapid quench. We hope that this work motivates further study of droplet tilings composed of two-component droplets. 

\section{Materials and methods}

\subsection{Experimental set-up}
We study the collective behavior of evaporating two-component droplets subject to an external field. The evaporating droplets are composed of propylene glycol (PG) and water, where the concentration C is denoted by percentage of PG over total volume. This particular concentration was chosen  because it corresponds to the region with the greatest stability in apparent contact angle for a wide range of relative humidities \cite{cira2015, benusiglio2018}. Additionally, droplets with a higher proportion of water (C < 10\%) evaporate too quickly, resulting in significant changes in their properties limiting the length of an experiment. A small amount of UV active dye is added to image the droplets under UV illumination. The amount of dye added is small enough where ternary effects can be ignored. Droplets are deposited in parallel following the method developed by our lab and described elsewhere \cite{molina2021}. Droplets are placed such that each is separated from one another by lattice constant a. In this work, we perform two sets of experiments where droplets are characterized by an average radius of $R = 2.75$ or $3.37 mm$, corresponding to $0.18a$ and $0.21a$, respectively. Droplets are confined to a hexagonal honeycomb lattice with a=7.5 mm and N=61 unit cells. Neighboring unit cells are separated by hydrophobic boundaries composed of gold treated with dodecane thiol of width w = 1.0 mm. The system is placed in an enclosed chamber and allowed to evolve. Previous work has shown that these conditions lead to a system dominated by diffusion where the length scale of droplet-droplet interactions depends on the density of the system.

Experiments were carried out in an enclosed chamber that minimized any disturbances to the droplets from ambient air currents. The droplet array is illuminated at a shallow angle with UV light ($\lambda = 400$ nm) and imaged at 1 FPS from a USB camera (See3CAM 130, eCon Systems) with a long pass filter cube ($\lambda = 495$ nm) placed above the lattice. A DHT-22 sensor (Adafruit Industries, USA) is placed in the chamber to measure the relative humidity ($57\% \pm 3\%$) and temperature ($24^{\circ}C \pm 1^{\circ}C$) at the boundary of the system. We note that the illumination did not lead to any significant increase in temperature during experiment and variations in humidity are small. This chamber sits atop a custom-built, rotary-actuated Stewart platform with 6 degrees of freedom \cite{Stewart1965, patel2018}. This design was chosen for its ability to support a large chamber with a high level of stability. Briefly, the set up consists of 6 servo motors (Dynamixel AX-12A, Robotis) controlled by an Arduino compatibile microcontroller (Arbotix-M Robocontroller, Trossen Robotics). The servos labelled $i=1-6$ and are arranged in 3 pairs separated by $120^{\circ}C$.

To coordinate the motion of the 6 servo motors so as to produce a precessing gravity vector, each of the $n$ pairs is given a phase offset of $2n\pi/3$.  The motion of the platform can then be controlled with a specified amplitude $A$ and period $T$ to give an angle update $\beta$

\begin{equation}
    \beta =A \cos{(\frac{2 \pi t}{T} + \frac{2n\pi}{3})}
\end{equation}

\noindent
This update is converted into microseconds and sent to the $i^{th}$ motor which is incremented depending on its location:

\begin{equation}
    \beta^{(i)}_{t+1} =
    \begin{cases} 
      \beta^{i}_{t} + \beta  & i, even \\
      \beta^{i}_{t} - \beta  & i, odd
   \end{cases}
\end{equation}

\noindent
The orientation of the platform is measured in two ways. First a bulls-eye spirit level is used as a check to ensure that the platform is level for static experiments. Second, a BNO-055 digital accelerometer (Bosch Sensortec GmbH, Germany) is used to measure the time evolution of the field during experiments. The sensor is configured to report quaternion values $(w, x, y, z)$ which can be used to construct a three dimension rotation matrix for every time step: 

\begin{equation}
    R = \quad
    \begin{pmatrix}
    w^2+x^2+y^2+z^2 & 2(xy-wz) & 2(wy+xz)\\
    2(xy+wz) & w^2-x^2+y^2-z^2 & 2(-wx+yz)\\
    2(-wy+xz) & 2(wx+yz) & w^2-x^2-y^2+z^2
    \end{pmatrix}
    \quad
\end{equation}

\noindent
The matrix $R$ can be used to obtain an orientation vector $\textbf{v}$ by through multiplication of the unit vector normal $\textbf{n} =  [0, 0, 1]^T$ to the platform surface in a flat configuration. The tilt angle $\alpha$ between this normal vector and the orientation vector is calculated according to:

\begin{equation}
    \alpha = \arccos \left(
    \frac
    {\textbf{v}\cdot\textbf{n}}
    {\| \textbf{v} \| \| \textbf{n} \|}
    \right)
\end{equation}

\noindent
Two dimensional projections of the field are obtained by taking the $x,y$ components of $\textbf{v}$ and converted to an angle $\theta=\arctan(y/x)$. The input function $u$ is defined as the cosine of $\theta$ relative to the unit vector $\textbf{n} =  [0, 1]^T$ and can take values between -1 and 1.   
 
\subsection{Numerical simulations}

The dynamics are obtained by integration of eqn. \ref{eqn:eqn_combined} with the addition of a third term $\nabla E_{cell}$ to represent the short ranged repulsive force imposed by the hydrophobic material used to fabricate the lattice pattern. Mathematically, this is represented as a logistic function whose parameters are chosen such that the barrier height is approximately an order of magnitude greater than either the terms $F_{g}$ or $F_{a}$:

\begin{equation}
    E_{cell} = \frac{-R}{1+e^{\frac{0.48-R-z(x,y)}{0.023}}}
\end{equation}

\noindent
Here, $z(x,y)$ reflects the unit cells hexagonal symmetry and determines the influence of the boundary given droplets Cartesian coordinates $(x,y)$:

\begin{equation}
    z = \frac{\sqrt{x^2 + y^2}}{c_{0} + \sum^{4}_{n}c_n\cos{(6n \kappa)}}
\end{equation}

\noindent
where $\kappa = \arctan{\frac{y}{x}}$ the denominator is a Fourier expansion in powers of 6 with coefficients $ c_{n} $ chosen such that derivatives with respect to $ \zeta $ are set equal to zero. The sharp boundary that this function creates makes this a system of stiff differential equations and so they are integrated using a backwards difference formula implemented by the NDSolve function in Mathematica. The rotating gravitational vector $u$ is modeled as $[\cos(2 \pi \omega t) \hat{x}, \sin(2 \pi \omega t) \hat{y}]$

\subsection{Data processing}
After each experiment, the data are offloaded from the camera and are first preprocessed using ImageJ before extracting droplet coordinates using a custom-built, Mathematica-based, data-processing pipeline. ImageJ is used to crop unnecessary data, rotate the frame to a standard orientation, and finally to perform image stabilization using the Lucas-Kanade algorithm \cite{ImageStabilizer}. The preprocessed images are then binarized by thresholding the green color channel with a binarization threshold value of 0.58-0.62. The $x,y$ coordinates of individual droplets are then tracked by assigning each droplet to a region defined by a mask describing the lattice geometry. The inbuilt Mathematica functions SelectComponents and MeasureComponents are used to obtain droplet coordinates and radii, respectively.

\subsubsection{Droplet orientation}

The orientation $\psi_{i}$ of individual droplets are obtained from $x,y$ coordinates according to $\psi_{i} = \arctan( y_{i} / x_{i} )$. The magnetization is given by $M = \frac{1}{N} \sum_{i}^{N} s_{i}$, where $s_{i}$ is the unit vector with the orientation $\psi_{i}$ of the $i^{th}$ droplet. The choice to define the state of each droplet to a unit vector was made to eliminate the effect of varying droplet size. $M$ can take values between 0 and 1. The collective phase response is given by $\Psi = \frac{1}{N} \sum_{i}^{N} \psi_{i}$. $\Psi$ is defined relative to the unit vector $\textbf{n} =  [0, 1]^T$ and can take values between -1 and 1. 

\subsubsection{Equal-time correlaton function}

The equal-time correlation function $C(\xi, t) = <s(r,t)s(r+\xi,t>$ is defined in terms of the angle formed between neighboring spin orientations that are $\xi$ lattice sites away. The angle between neighboring spins is calualted using the law of cosines. The characteristic blob size $L$ is defined as the length scale at which $C(\xi = L) =1/2$ \cite{sethna2006}.

\subsubsection{Specific heat}

The specific heat $<M^{2}> - <M^{2}>$ is defined as the time-averaged fluctuations in magnetization. The spatial dependence of $C_{v}$ was characterized by calculating $M$ from a subset of droplets. Specifically, $M = \frac{1}{\mathcal{N}(\xi)} \sum^{\mathcal{N}(\xi)}_{i} s_{i}$, where $\mathcal{N}(\xi)$ is the set of droplets that are $\xi$ lattice sites from the center of the system. 

\subsubsection{Bond lengths}
Bond lengths are calculates bby taking the difference between the $x,y$ coordinates that of droplets that are $\xi=1$ lattice sites away and normalized by the time-averaged bond length for that system. 

\subsubsection{Vertex counts}
Vertex counts are determined by the number of droplets located $0.15a$ ($2.4$ $mm$) from a given vertex site. A droplet is assigned to a given vertex if its center is within this range.

\section*{Author Contributions}

AM is credited with investigation, visaluzation of data, and writing the original manuscript. AM, MP are credited with conceptualization, formal analysis, and reviwing \& editing the manuscript. MP is credited with funding acquisition and supervising the research. 

\section*{Conflicts of interest}
There are no conflicts to declare.

\section*{Acknowledgements}
We thank all members of PrakashLab for valuable discussions - particularly we would like to acknowledge Rahul Chajwa for thoughtful discussions and providing commentary on the manuscript; Rebecca Konnte and Daniel Sheykavich for art and photography; Stefan Karpitschka for contributing to the early development of the numerical simulations and early experimental design; Anna Lei for contributing to early experimental design. AM received funding from the National Science Foundation Graduate Research Fellowhsip Program. MP acknowledges support from the Keck Foundation, Moore foundation, and Schmidt innovation Fellow Award.

\newpage
\FloatBarrier
\bibliographystyle{naturemag}
\bibliography{main}

\end{document}


\maketitle

\noindent
\author{Anton Molina$^{a, b}$, Manu Prakash$^{b, \ast}$

\noindent
\normalsize{$^{a}$Department of Materials Science and Engineering, $^{b}$Department of Bioengineering}

\noindent
\normalsize{Stanford University, 450 Serra Mall, Stanford, California 94305}

\noindent
\normalsize{\affil[$^\ast$] To whom correspondence should be addressed; E-mail:  manup@stanford.edu}\\

This supplementary file includes:

\begin{itemize}
\setlength\itemsep{-0.75em}
    \item Supplementary Figures 1-8
    \item Supplementary Videos 1-6
\end{itemize}

\newpage
\FloatBarrier
\begin{figure}[h]
    \centering
    \includegraphics[scale=0.9,width=0.9\textwidth]{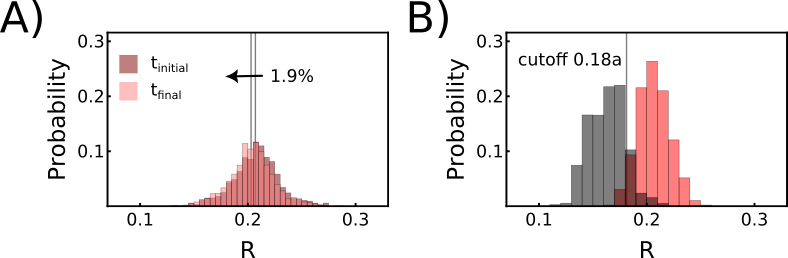}
    \renewcommand{\figurename}{Supplementary Figure}
    \renewcommand{\thefigure}{1}
    \caption{Size distribution of droplets. (A) distribution of droplets used to obtain experimental phase diagram in Fig. 2. Measurement of droplet radius at $t_{initial}$ and $t_{final}$ show $1.9\%$ reduction in size. (B) distribution for the two systems represented in Fig. 4. The cutoff described in the text is indicated by the vertical line.}
    \label{fig:SI_fig1}
\end{figure}

\newpage
\FloatBarrier
\begin{figure}[h]
    \centering
    \includegraphics[scale=0.9,width=0.9\textwidth]{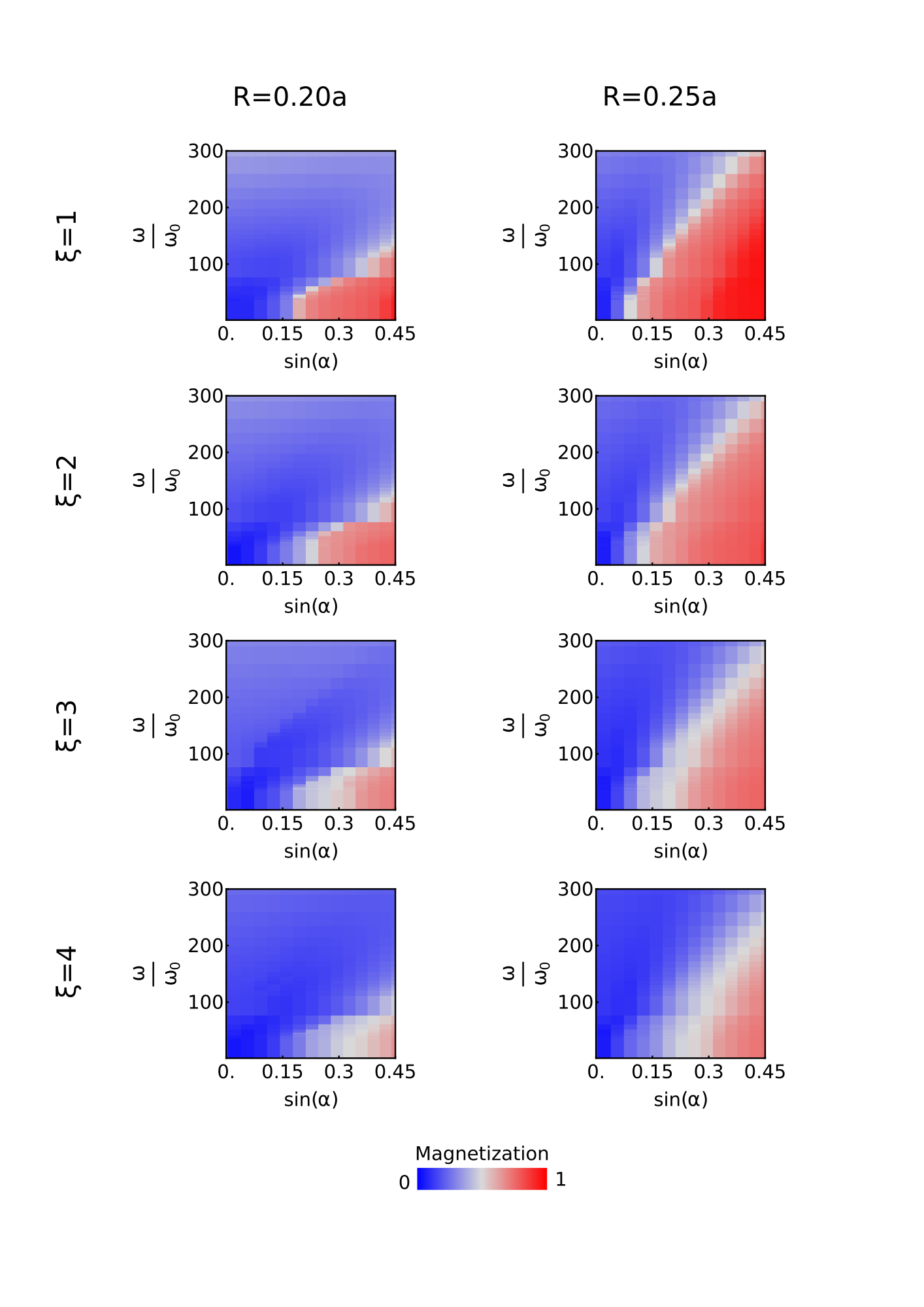}
    \renewcommand{\figurename}{Supplementary Figure}
    \renewcommand{\thefigure}{2}
    \caption{Numerical phase diagrams for uniform droplet sizes and varying range of interactions $\xi$.}
    \label{fig:SI_fig2}
\end{figure}

\newpage
\begin{figure}[h]
    \centering
    \includegraphics[scale=0.8,width=0.8\textwidth]{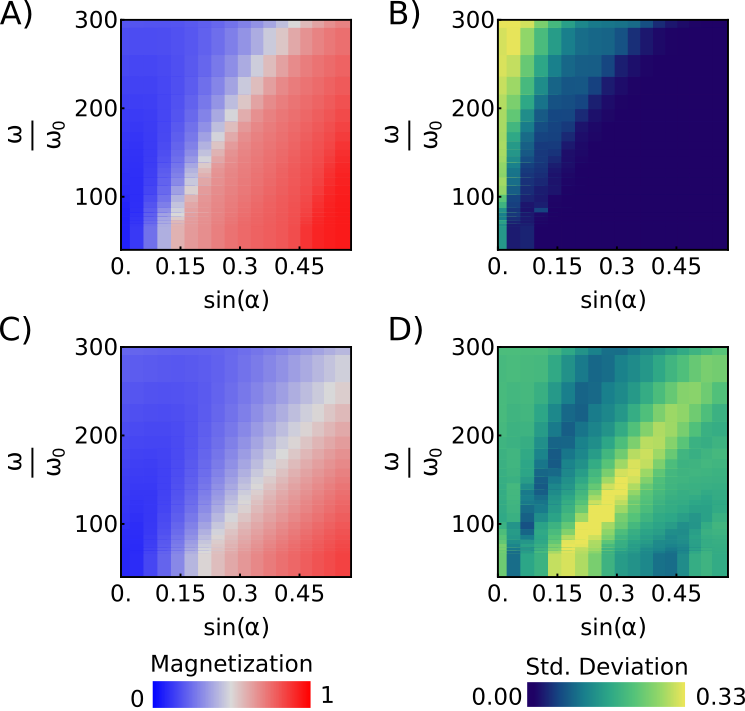}
    \renewcommand{\figurename}{Supplementary Figure}
    \renewcommand{\thefigure}{3}
    \caption{Effect of different sources of noise on numerical phase diagram. Numerical phase diagram (A) and variance (B) for uniform droplets ($R=0.25l$) with 5\% noise in initial position. Numerical phase diagram (C) and variance (D) for uniform droplets ($R=0.225l$) with 10\% noise in $R$. All simulations were fixed with $\xi=2$ and each diagram represents $N=6$ different random initial conditions.}
    \label{fig:SI_fig3}
\end{figure}

\newpage
\FloatBarrier
\begin{figure}[h]
    \centering
    \includegraphics[scale=0.8,width=0.8\textwidth]{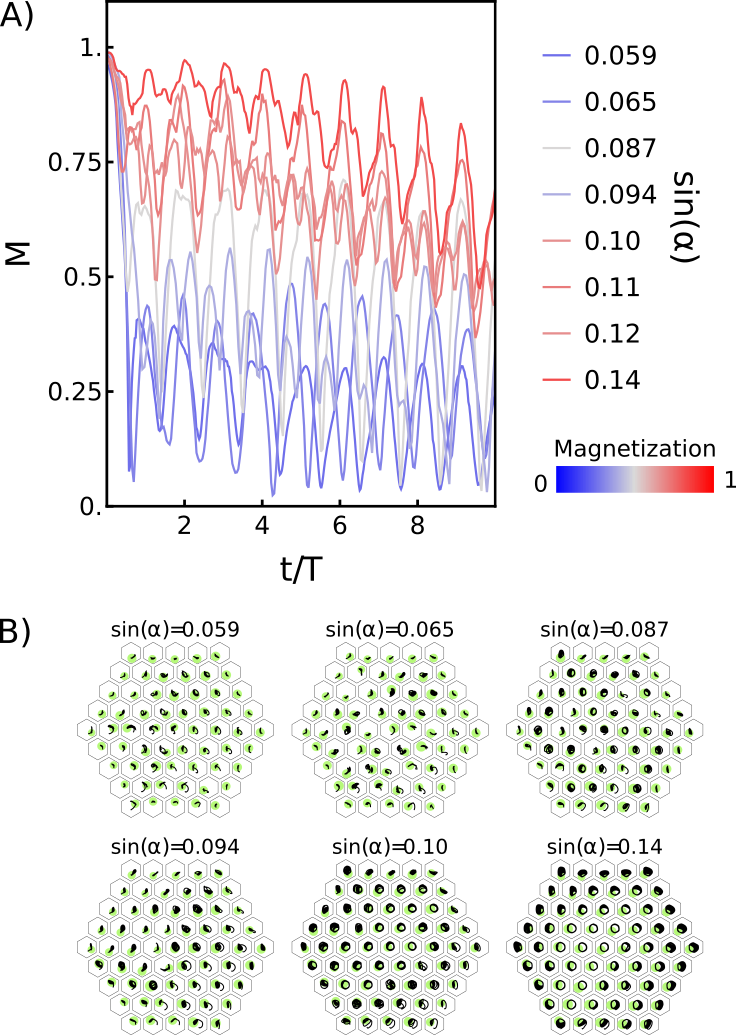}
    \renewcommand{\figurename}{Supplementary Figure}
    \renewcommand{\thefigure}{4}
    \caption{(A) Dynamics of order parameter for experimental realization ($R=0.18a$, noise=$10.1\%$).(B) Trajectories of individual droplets shown for values of $sin(\alpha)$ selected from (A).}
    \label{fig:SI_fig4}
\end{figure}

\newpage
\FloatBarrier
\begin{figure}[h]
    \centering
    \includegraphics[scale=1.0,width=1.0\textwidth]{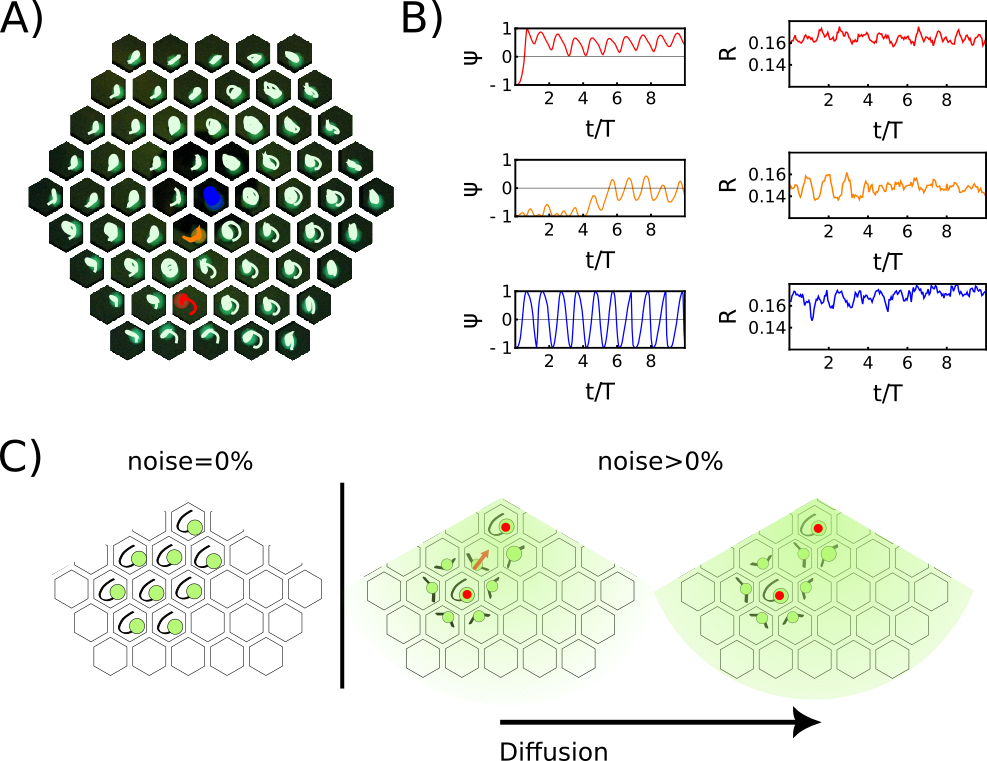}
    \renewcommand{\figurename}{Supplementary Figure}
    \renewcommand{\thefigure}{5}
    \caption{Microscopic dynamics in transition region depends on droplet size and local environment. (A) Droplet trajectories for an experiment conducted at $\sin(\alpha)=0.095$ and $\omega / \omega_{0} = 235$ (Fig. 4E) with three droplets highlighted. (B) Time evolution of $\psi$ (left) and radius $R$ (right) for the highlighted droplets in (A) corresponding to fixed (red), migrating (orange), and field-driven (blue) droplets. (C) Cartoon illustration showing possible scenario where diffusion and details of neighboring droplets work together to create migratory droplet behavior. Migrating droplets are an example of non-trivial dynamics. }
    \label{fig:SI_fig5}
\end{figure}

\newpage
\FloatBarrier
\begin{figure}[h]
    \centering
    \includegraphics[scale=0.8,width=0.8\textwidth]{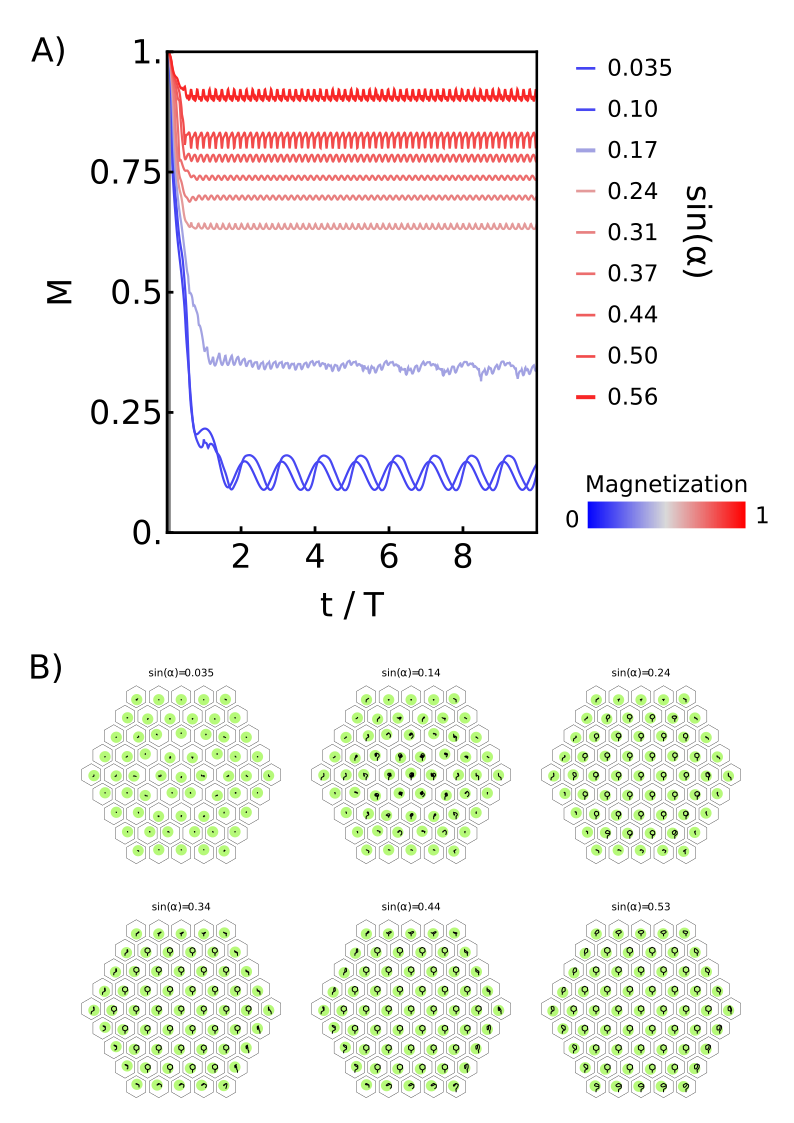}
    \renewcommand{\figurename}{Supplementary Figure}
    \renewcommand{\thefigure}{6}
    \caption{(A) Dynamics of order parameter for uniform droplets ($R=0.25l$, noise=$0\%$). (B) Trajectories of individual droplets shown for values of $sin(\alpha)$ selected from (A).}
    \label{fig:SI_fig6}
\end{figure}

\newpage
\FloatBarrier
\begin{figure}[h]
    \centering
    \includegraphics[scale=0.8,width=0.8\textwidth]{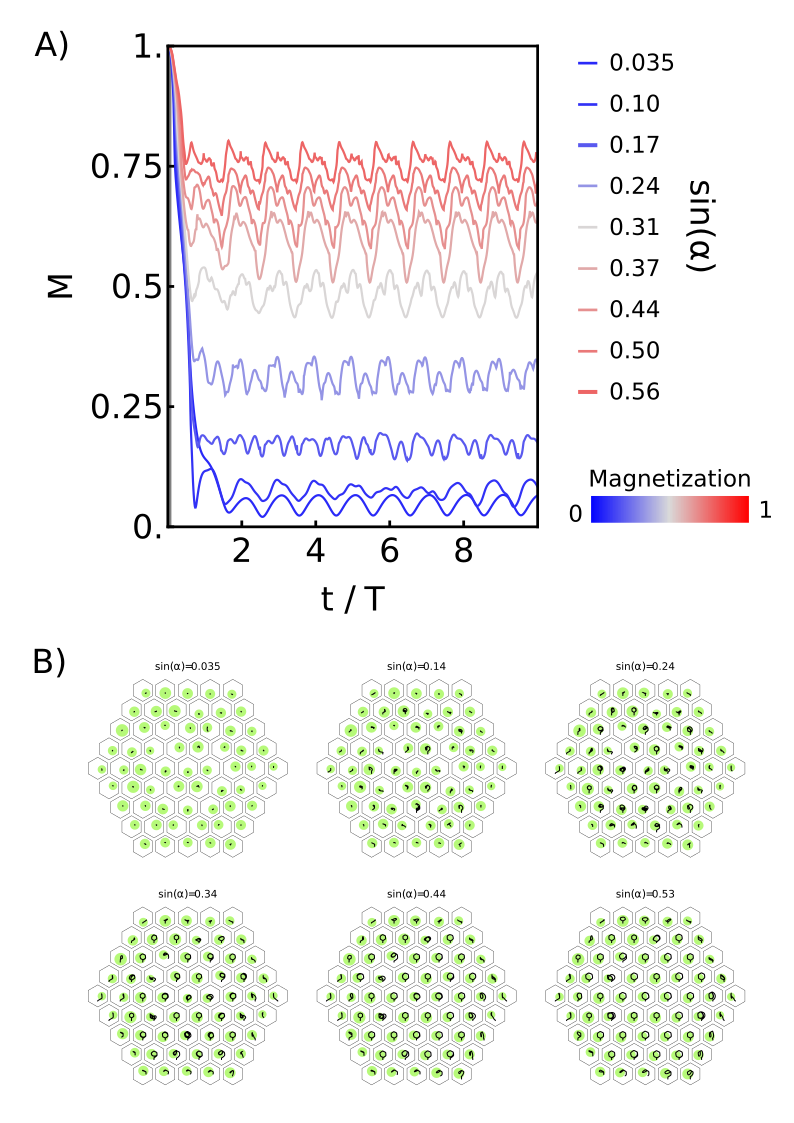}
    \renewcommand{\figurename}{Supplementary Figure}
    \renewcommand{\thefigure}{7}
    \caption{(A) Dynamics of order parameter for disordered droplets ($R=0.225a$, noise=$10\%$). (B) Trajectories of individual droplets shown for values of $sin(\alpha)$ selected from (A).}
    \label{fig:SI_fig7}
\end{figure}

\newpage
\FloatBarrier
\begin{figure}[h]
    \centering
    \includegraphics[scale=0.9,width=0.9\textwidth]{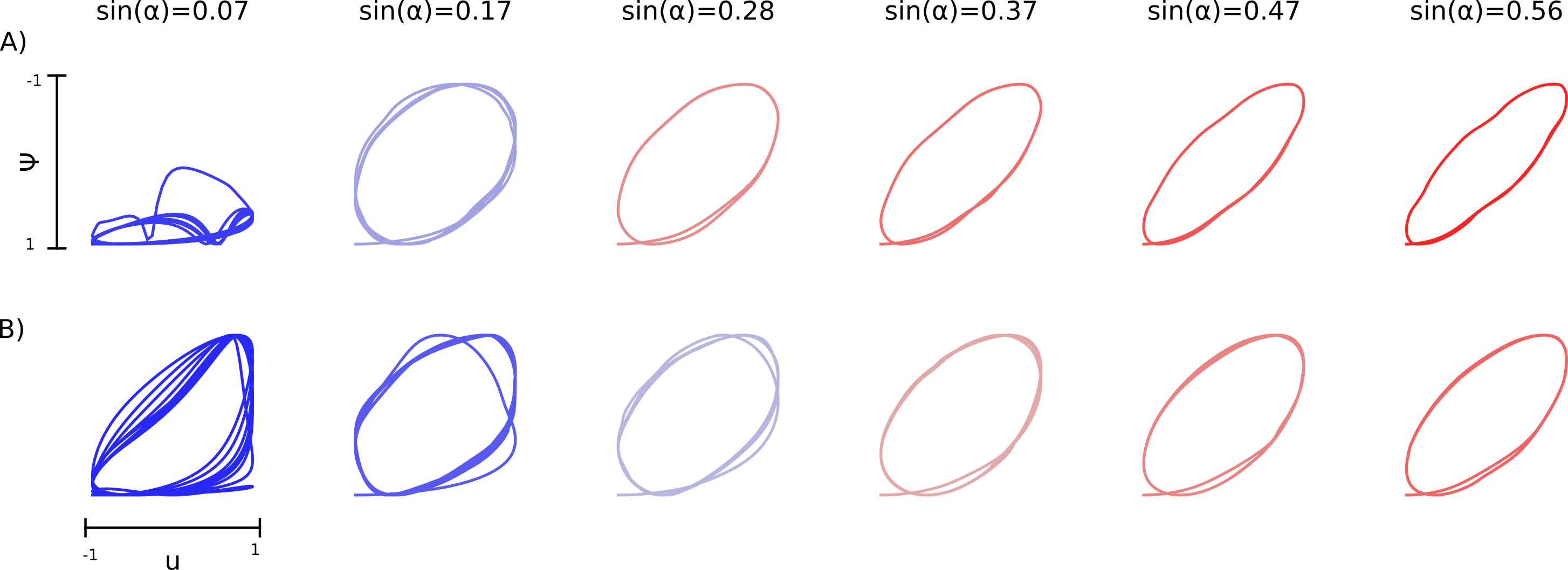}
    \renewcommand{\figurename}{Supplementary Figure}
    \renewcommand{\thefigure}{8}
    \caption{Hysteresis loops obtained from simulation data at ($\omega / \omega_{0} = 160$) with (A) uniform droplet size ($R=0.25a$, noise$=0$) and (B) disordered droplet size ($R=0.225a$, noise$=10\%$). For each loop the horizontal axis is the orientation of the applied field and the vertical axis is magnetization. The loops are colored according to their average magnetization according to the color scheme described in Fig. 2.}
    \label{fig:SI_fig8}
\end{figure}


\newpage

\noindent
\textbf{Supplementary Video 1:} Experimental set showing preparation of a droplet tiling, placement in an enclosed chamber, and implementation of rotating gravitational field using a Stewart platform. 

\noindent
\textbf{Supplementary Video 2:} Transition from interaction- to field-dominated behavior controlled by strength of rotating gravitational field. Representative data for three distinct phases are shown. The orientation of droplets are annotated according the scheme in Fig. 1. 

\noindent
\textbf{Supplementary Video 3:} Numerical simulation showing the transition from interaction-dominated to field-dominated behavior at $\omega / \omega_{0} = 160$ for a system with $R=0.225a$ and noise=$10\%$. The interaction lengthscale is set to $\xi=2$

\noindent
\textbf{Supplementary Video 4:} Mesoscale structures emerge at the onset of transition region. The qualitative features of the structures depend on the spatial distribution of field-couplings. The orientation of droplets are annotated according to the scheme in Fig. 1. Bond Lengths between neighboring droplets are annotated according to the scheme in Fig. 4.

\noindent
\textbf{Supplementary Video 5:} Strongly, driven large droplets leave behind a trail of their own constituents. 

\noindent
\textbf{Supplementary Video 6:} Annealing of the frustrated centered hexagonal honeycomb lattice ($N=61)$. The orientation of droplets are annotated according to the scheme in Fig. 1. Bond Lengths between neighboring droplets are annotated according to the scheme in Fig. 4.